# Correctly validating results from single molecule data: the case of stretched exponential decay in the catalytic activity of single lipase B molecules

Ophir Flomenbom[1,§], Johan Hofkens[2,§], Kelly Velonia[3],

Frans C. de Schryver[2], Alan E. Rowan[4,§], Roeland J. M. Nolte[4],

Joseph Klafter[5], Robert J. Silbey[1]

[1] Department of Chemistry, Massachusetts Institute of Technology, Cambridge, MA 02139, USA

[2] Department of Chemistry, Katholieke Universiteit Leuven, Celestijnenlaan 200 F, 3001 Heverlee, Belgium

[3] Department of Organic Chemistry, University of Geneva, 30 quai Ernest Ansermet, Geneva 4, CH-1211, Switzerland

[4] Department of Organic Chemistry, Radboud University Nijmegen, 6525 ED Nijmegen, The Netherlands

[5] School of Chemistry, Raymond & Beverly Sackler Faculty of Exact Sciences, Tel Aviv University, Ramat Aviv, Tel Aviv 69978, Israel

[§] Corresponding Authors; J.H: email address: johan.hofkens@chem.kuleuven.be fax: +32 (0)16 327989, O.F: email address: flomenbo@mit.edu fax: 1+617-253-7030, A.E.R email address: A.Rowan@science.ru.nl fax: +31 (0)24 365 29 29.

**Abstract**

The question of how to validate and interpret correctly the waiting time probability density functions (WT-PDFs) from single molecule data is addressed. It is shown by simulation that when a stretched exponential WT-PDF, $\phi_{off}(t) = \phi_0 e^{-(t/\tau)^\alpha}$, generates the *off* periods of a two-state trajectory, a reliable recovery of the input $\phi_{off}(t)$ from the trajectory is obtained even when the bin size used to define the trajectory, *dt*, is much larger than the parameter $\tau$. This holds true as long as the first moment of the WT-PDF is much larger than *dt*. Our results validate the results in an earlier study of the activity of single Lipase B molecules and disprove recent related critique.

**Introduction**

Advances in single molecule experimental techniques allow studying at room temperature the dynamics of a large number of processes in real time. Examples include the flux of ions through individual channels [1], conformational fluctuations of biopolymers [2-5], and kinetics of single enzymes [6-9]. The aim of such sophisticated measurements is to learn about the underlying mechanism of the process to an extent that is unattainable from bulk measurements due to averaging.

In the simple scenario, the first step in the analysis of the data turns it into a trajectory of *on* and *off* periods (waiting times), FIG 1A. A frequently used assumption describes the mechanism of the observed process by a multi-substate *on-off* Markovian kinetic scheme (KS) [10-17], Fig. 1B; see also [18-21] for other model types. Using statistical tools of data analysis [22-26], one wishes to learn as much as possible about the underlying KS.

It was recently shown in [10], that the way to utilize the information content in the two-state trajectory is to construct from it a canonical form of reduced dimensions. The mathematical mapping from the space of KSs into the space of reduced dimension forms enables finding, in the most efficient way, the relationship between the data and all possible KSs that can generate it. Technically, the analysis uses, along with other methods, the two dimensional histograms of successive events.

However, there are cases in which the experimental two-dimensional histograms are too noisy due to the limited number of events in the data, and other methods of analysis are used to detect correlations between successive events. This was the case in the study of the activity of single lipase B molecules from *Candida antartica* [8-9]. The two-state trajectories contained approximately $10^4$ cycles. The bin size of the trajectories was 1*ms*. The first stage in the analysis of the two-state trajectories constructed the WT-PDFs of the *on* and *off* periods. (This way of analysis is the most basic one when analyzing two-state trajectories, and must be accompanied by methods that look for correlations between successive events [8-9].) The *on* WT-PDF was approximated by an exponential decay with a fast decaying rate of the order of the bin size, which reflects short *on* periods. In contrast, the *off* periods were, on average, two orders of magnitude larger than the bin size. The experimental histograms for the *off* periods, at saturating substrate concentrations, were fitted by a stretched exponential, $\phi_{off}(t) = \phi_0 e^{-(t/\tau)^\alpha}$, with the parameters $\alpha = 0.15$ and $\tau = 0.00115 ms$, independent of the substrate concentration. The limited number of events in the trajectory dictates analyzing the *off* ordered waiting times trajectory, which is the trajectory of the *off* waiting times plotted vertically as a function of the occurrence index. This trajectory revealed correlations between successive *off* periods. Combining these two features, along with other experimental findings, led to a mechanism that involves simultaneously conformational changes and enzymatic activity (Fig. 1B). Importantly, the mechanism strongly deviates from the Michaelis-Menten mechanism that has been used to model enzymatic activity for almost a century, see for example [27].

A recent Letter [28], incorrectly argues against the interpretation of the experimental *off* histogram found in Refs. [8-9] as a stretched exponential *off* WT-PDF. The critique in Ref. [28] wrongly concludes that the values for $\alpha$ and $\tau$ found for the experimentally stretched exponential *off* histogram cannot be extracted from a two-state trajectory with a bin size of 1*ms* whose *off* periods were generated from such a WT-PDF. By means of a simulation, it is shown that a stretched exponential WT-PDF with the parameters $\alpha = 0.15$ and $\tau = 0.00115 ms$ is reliably recovered from a $10^4$-cycle two-state trajectory with a bin size of 1*ms*, whose *off* periods are generated by this WT-PDF. This result holds true as long as the first moment of the WT-PDF is still much larger than the bin size. We also point out on the technical error that led to the incorrect conclusion in [28]; the wrong choice of the prefactor in the plot of the experimental WT-PDFs led to the wrong conclusion in [28].

**2 – How to interpret experimental results correctly?**

*2.1 - Moments analysis* Consider the normalized WT-PDF,

$$\phi_{off}(t) = \phi_0 e^{-(t/\tau)^\alpha} \equiv \phi_0 \psi_{off}(t) \quad ; \quad \phi_0 = 1/\int_0^\infty \psi_{off}(t)dt = \frac{\alpha/\tau}{\Gamma(1/\alpha)}. \quad (1)$$

The $n^{th}$ moment of the WT-PDF in Eq. (1) is given by,

$$<t_{off}^n> = \int_0^\infty t^n \phi_{off}(t)dt = \tau^n \frac{\Gamma[(n+1)/\alpha]}{\Gamma(1/\alpha)},$$

where $\Gamma(.)$ is the gamma function.

In particular, the first moment, *n=1*, with,

$\alpha = 0.15$, $\qquad\qquad \tau = 0.00115 ms$, $\qquad\qquad\qquad\qquad$ (2)

has the value of,

$<t_{off}> = 3300$ *ms*.

Although the functional form for the *off* WT-PDF in [8-9] was found to follow Eq. (1) with the parameters given by Eq. (2), the value of the average *off* waiting times found directly from the data was only 220 *ms*. The reason for this discrepancy is very simple; the largest random *off* period observed in the experiment was about 1000*ms*. Accordingly, when calculating the moments (and the normalization) of the WT-PDF found from the experiment using Eq. (1), one should take into consideration the time window relevant to the experiment. With a cutoff of 1000 *ms*, the numerically obtained first moment is consistent with the experimental value,

$<t_{off}> = 210 ms$ $\qquad ; \qquad t_{off,i} \leq 1000 ms$, $\forall i$ .

The above discussion emphasizes the subtleness of testing for self-consistency; one must correctly set the time window of the integration used in calculating the moment and the normalization of the WT-PDF, whose temporal decay pattern is found from the experiment, when comparing it with the average of the *off* periods found directly from the trajectory.

***2.2 Plotting analytical WT-PDFs and experimentally obtained WT-PDFs*** Suppose that the analytical WT-PDF in Eq. (1) with the parameter values in Eq. (2) is plotted with $dt=1ms$. In a plot of the ratio $\phi_{off}(t)/\phi_{off}(0)$, the normalization constant vanishes, and we end up plotting $\psi_{off}(t)$ (Fig. 2A). Thus, a simple manipulation of the axes immediately leads to a straight line for the above ratio. Namely, when plotting $f(\phi_{off}(t)/\phi_{off}(0))$ as a function of $\log t$, where $f(.)$ stands for a $-\log[-\log(.)]$ and $\log(.)$ is a natural $\log(.)$, we get a straight line for a normalized stretched exponential $\phi_{off}(t)$ (Figs. 2B). From this plot the extraction of the parameters $\alpha$ and $\tau$ is straightforward. Now, consider an experimental WT-PDFs constructed as a function of discrete time, $t=jdt,\ j=1,2,...,N$, where $Ndt$ is the largest relevant period. Note that the first value of $t$ equals to the bin size $dt$, which is larger than zero. So, if one looks at the ratio $\phi_{off}(t)/\phi_{off}(dt)$, which is equal to $\psi_{off}(t)/\psi_{off}(dt)$, and plots $f(\phi_{off}(t)/\phi_{off}(dt))$ versus $\log t$, one clearly does not get a straight line (Figs. 2C-2D). The deviation of this ratio from a straight line does not mean that $\psi_{off}(t)$ obeys a decay pattern different than a stretched exponential. It simply means that $\psi_{off}(dt) \neq 1$. Equation (1) implies, $\psi_{off}(dt) \leq 1$, where the equality holds only in the limit $dt \to 0$, but in experimental conditions $dt$ has a finite value. If one builds a histogram from an experiment as,

$\tilde{g}_{exp}(t) =$ # *of periods between t-dt and t / # of periods*,     *j=1,2,...,N,*

(where as in [8-9], only the right edge of the *j* bin, among its two edges, contributes to $\tilde{g}_{exp}(t)$ and the histogram bin size is taken to be equal to the trajectory bin size) and further looks at the ratio,

$g_{exp}(t) = \tilde{g}_{exp}(t)/\tilde{g}_{exp}(dt)$,

one ends up looking at (or estimating),

$g_{\exp}(t) = \psi_{off}(t) / \psi_{off}(dt)$.

To proceed correctly with the analysis, one should search for a constant $c$, such that plotting the product $cg_{\exp}(t) [\equiv G_{\exp}(t)]$ in the manipulated axes leads to a straight line. If such a constant exists, it is unique, and $G_{\exp}(t)$ approximates $\psi_{off}(t)$. This demand on $G_{\exp}(t)$ is equivalent to the condition, $(G_{\exp}(t))_{t \to 0} = 1$, which was required in [9]. (In fact, $c = \psi_{off}(dt)$. $c$ can be easily found from the first couple of data points.)

Note that the above means that for any value of $t$,

$$\frac{\psi_{off}(t)/\psi_{off}(dt)}{\phi_{off}(t)/\phi_{off}(0)} = e^{+(dt/\tau)^\alpha} \geq 1 ,$$

where the equality holds in the limit $dt \to 0$. This explains figures 1 and 2 in [28]; figure 1 in [28] plots curves with improper prefactors, and figure 2 in [28] plots $e^{+(dt/\tau)^\alpha}$ as a function $dt$ for different values of $\alpha$ and $\tau$. Thus, a general conclusion regarding the disability of reconstructing the WT-PDF in Eqs. (1)-(2) from a $1ms$ trajectory cannot be drawn from the analysis given by Molski [28]. Moreover, below it is shown by simulation that reconstructing the WT-PDF of Eqs. (1)-(2) with $1ms$ bin size is easily performed.

***2.3 - Self consistent test: drawing random times from a stretched exponential WT-PDF and constructing it back*** The simplest self-consistent test for reconstructing the stretched exponential WT-PDF in Eqs. (1)-(2) from a trajectory with $dt=1ms$ is performed. The numerical test draws random times from $\psi_{off}(t)$ using the rejection method [29]. Figure 3A plots $f[G_{\exp}(t)]$ obtained from a $10^4$-cycle trajectory, with $c = 0.063$. Also shown is $f[\psi_{off}(t)]$. These WT-PDFs coincide. Importantly, the decay of $G_{\exp}(t)$ spans two to three orders of magnitude both in time and in amplitude. For comparison, Fig. 3B plots the same curves, but now $G_{\exp}(t)$ (with the same value of $c$) is obtained from a $10^6$-cycle trajectory. As expected, $G_{\exp}(t)$ built from the longer trajectory approximates better the input WT-PDF. However, a trajectory with $10^4$ cycles is long enough for $G_{\exp}(t)$ to reliably approximate the temporal decay of the true WT-PDF, and allows an accurate parameter extraction.

## 3 – Concluding Remarks

In this Letter, it was shown by simulation that a reliable stretched exponential decay pattern with the parameters $\alpha = 0.15$ and $\tau = 0.00115 ms$ can be recovered from a two-state trajectory with *dt*=1*ms* whose *off* periods are generated by this WT-PDF. This validates the interpretation of the experimental findings made in [8-9] that for saturating substrate concentrations the *off* periods in the catalytic activity of single lipase B molecules from *Candida antartica* construct a stretched exponential WT-PDF, and disproves the critique made in [28] about the correctness of such an interpretation. The result of this numerical test should not be a surprise because the moments of the WT-PDF in Eq. (1) with the parameters in Eq. (2) are all much larger than the bin size of 1*ms*. The basic physical relevant timescale of a WT-PDF is its first moment (if it is finite). The misinterpretation of $\tau$ as the timescale in the problem may arise from inappropriate comparison between a stretched exponential WT-PDF and an exponential WT-PDF. A stretched exponential WT-PDF is characterized by 2 parameters, $\alpha$ and $\tau$. Strictly speaking, only for $\alpha = 1$ is $\tau$ the relevant timescale in the problem. Although one can consider $\tau$ as the relevant timescale in the problem even in the vicinity $\alpha \approx 1$, in the case of Refs. [8-9], the small value of $\alpha\,(=0.15)$ means a highly non-exponential decay pattern. This, in turn, assigns an important role for $\alpha$ in the determination of the timescale in the problem, which is evident in the moment expressions.

An explanation for the error that led to the wrong critique in [28] is also given. The use of an incorrect prefactor when plotting the experimental curves in the manipulated axes led to the wrong conclusion. It should be noted that for a stretched exponential WT-PDF with a first moment comparable or smaller than the bin size, difficulties in a reliable recovery of the WT-PDF from the trajectory are expected to arise. Such a situation can happen with any form of a WT-PDF; it is a general outcome of the loss of information about the measured process when extracted from data with 'bad' time resolution.

The occurrence of a stretched exponential decay pattern in measurements of protein activity has a long history, e.g. [30]. The origin for this behavior is usually explained by an ensemble of states, or conformations, that contribute to the activity in a particular way. This argument was used in [8-9] to explain the observed stretched exponential decay pattern. Moreover, the findings in [8-9] showed not just that the periods of the catalytic activity in between successive dissociation events of product molecules from the enzyme construct a highly stretched exponential WT-PDF, but also that successive such periods are correlated. The implementation of the stretched exponential decay and correlations in the observed catalytic activity into a mechanistic description led to the KS in Fig. 1B, and can be viewed as constrains on, or, properties of, the underlying energy landscape of the process. These properties of the energy landscape can be further related to properties deduced from (e.g.) a theoretical study of the process. Establishing such relationships will most certainly constitute an important step forward in the understanding of biopolymer activity.

**Figure Captions**

FIG 1A A trajectory of an observable that fluctuates between two values, *on* and *off*, as a function of time. Such a two-state trajectory can be generated from kinetic Monte-Carlo simulations, which model random walks on kinetic schemes, or on reduced dimension forms [10].

FIG 1B An example for a KS with only reversible transitions. The *off*-state consists of a spectrum of coupled *off* substates (black circles), whereas the *on* substates (red squares) are uncoupled. Each *on* substate is connected to a single *off* substate, and vice versa. This KS was used in [8-9] as a model for the enzymatic activity of lipase B from Candida Antartica.

Fig 2 $\psi_{off}(t)$ (**A**), $f[\psi_{off}(t)]$ (**B**), $\psi_{off}(t)/\psi_{off}(dt)$ (**C**) and $f[\psi_{off}(t)/\psi_{off}(dt)]$ (**D**) as a function of log(*t*). Although $f[\psi_{off}(t)]$ follows a straight line when plotted against log(*t*) (**B**), the ratio $f[\psi_{off}(t)/\psi_{off}(dt)]$ does not (**D**). This can explain the results in [28], where figure 1 in [28] plots curves with similar prefactors as in **D**. Here, and in the next Fig., $f[g(t)]$ stands for, -log(-log[$g(t)$]).

Fig 3 $G_{exp}(t)$ (dotted curve) with $c = 0.063$ and $\psi_{off}(t)$ (circled symbol) in the manipulated axes. In **A**, $10^4$ random times were used to construct $G_{exp}(t)$ where in **B** $10^6$ random times were used to construct $G_{exp}(t)$. Clearly, a $10^4$-cycle trajectory allows a reliable construction of $G_{exp}(t)$ and an accurate parameters extraction.

# Figures

FIG. 1

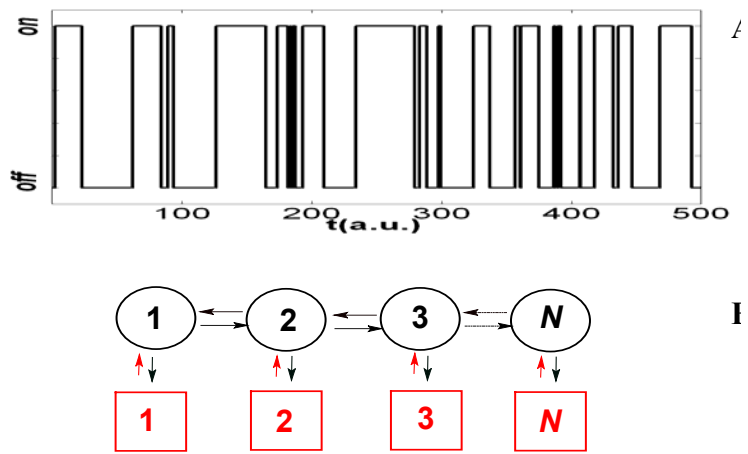

FIG. 2

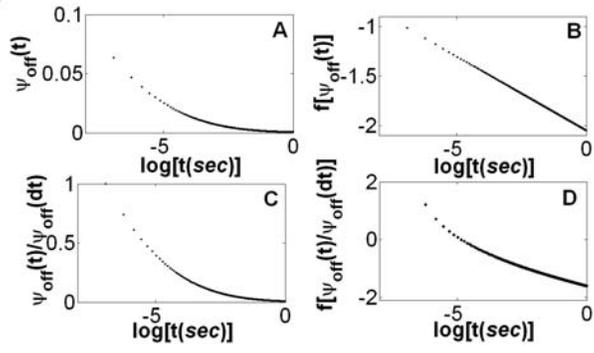

FIG. 3

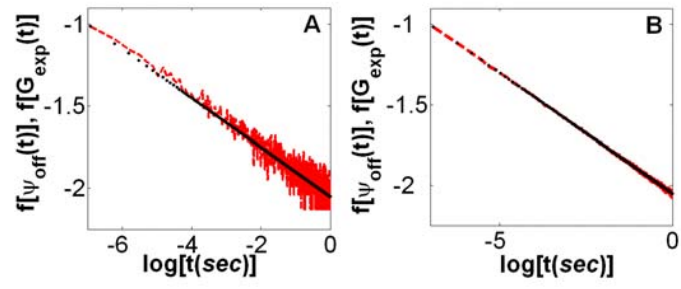